\documentclass[11pt]{article}

\usepackage[final]{acl}
\usepackage{booktabs}
\usepackage{multirow}
\usepackage{graphicx}
\usepackage{subcaption}
\usepackage{times}
\usepackage{latexsym}

\usepackage[T1]{fontenc}

\usepackage[utf8]{inputenc}

\usepackage{microtype}

\usepackage{inconsolata}
\usepackage[ruled,linesnumbered,lined,boxed,commentsnumbered]{algorithm2e}
\usepackage{amsmath}
\usepackage{amssymb}
\usepackage{colortbl}
\usepackage{float}
\usepackage{xcolor}
\usepackage[most]{tcolorbox}
\usepackage{environ}
\usepackage{tabularx}

\title{SALT: Salience-Aware Lexical Trie for Long-Context Compression}

\author{
  Oteo Mamo\thanks{Equal contribution.}\thanks{Corresponding author.} \quad
  Hyunjin Yi\footnotemark[1] \quad
  Joydhriti Choudhury \quad
  Shangqian Gao \quad
  Weikuan Yu \\
  Department of Computer Science, Florida State University \\
  Tallahassee, FL, USA \\
  \texttt{\{om21d, hy22c, jc23bc, sg24bi, wyu3\}@fsu.edu}
}

\begin{document}

\maketitle

\begin{abstract}
As large language models (LLMs) process increasingly longer prompts, computation and KV-cache memory costs have emerged as major bottlenecks in inference systems. Existing input-level prompt compression methods address this, but rank each sentence by a scalar relevance score, treating the document as an unstructured pool of words and sentences. Under tight budgets, this causes theme collapse, where the dominant themes of a document consume the budget, discarding less-frequent yet task-relevant themes. Preserving thematic coverage instead requires allocating the budget across recurring themes rather than scoring sentences in isolation. To this end, we propose SALT, a model-agnostic extractive framework that organizes per-sentence keywords into a trie ordered by sentence frequency (SF), a lightweight, reusable proxy for document thematic structure. This trie-based organization smooths memory allocation and prevents dominant themes from monopolizing the budget. Multi-anchor retrieval activates trie nodes labeled by query keywords at any depth, and the trie persists across dialogue turns, supporting multi-turn use without re-encoding the document. By preserving document themes, SALT reduces the prefill computation and memory cost of long-context prompts while remaining composable with KV-cache methods that target decoding-time latency and memory.
\end{abstract}

\section{Introduction}

Large language models (LLMs) have become essential for a wide range of natural language tasks, and their utility increasingly depends on their ability to process long-context inputs containing tens of thousands of tokens. These long contexts are central to document summarization, multi-hop question answering, and retrieval-augmented generation (RAG), where relevant information often spans long documents or multiple sources. To meet this need, recent open and proprietary models have expanded their context windows beyond a hundred thousand tokens \citep{dubey2024llama3herdmodels}. This expansion has unlocked new capabilities, but it has also exacerbated the computational and memory costs of long-context inference. Each additional input token increases prefill computation, expands the key-value (KV) cache held in GPU memory throughout inference, and spreads attention across more positions \citep{dao2024flashattention2}. As a result, latency, memory usage, and generation accuracy become increasingly constrained by input length, making prompt compression or reduction before inference one of the most direct ways to improve long-context efficiency without modifying the underlying model.

\begin{figure}[t]
  \centering
  \begin{subfigure}{0.48\columnwidth}
    \includegraphics[width=\columnwidth]{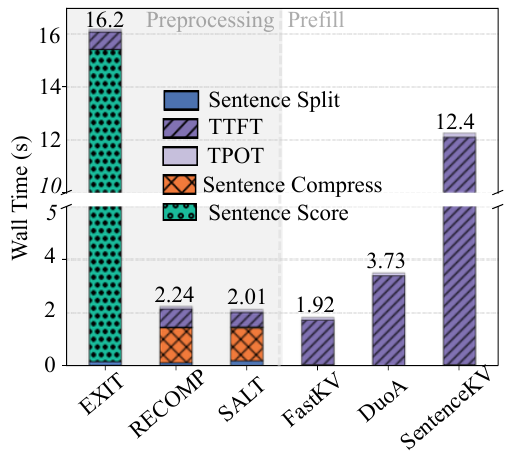}
    \caption{32k context length.}
    \label{fig:TTFT_comparison_30}
  \end{subfigure}
  \hspace{0.1pt}
  \begin{subfigure}{0.48\columnwidth}
    \includegraphics[width=\columnwidth]{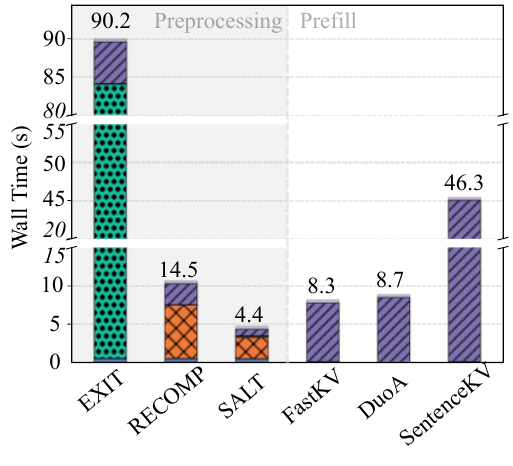}
    \caption{128k context length.}
    \label{fig:TTFT_comparison_100}
  \end{subfigure}
    \caption{Wall-time for preprocessing methods (EXIT, RECOMP, SALT) and prefill/KV cache methods (FastKV, DuoA, SentenceKV), averaged over 30 inputs at 32k and 128k context lengths. Bars show TTFT breakdown and averaged 64-token decoding as TPOT. All methods run with 20\% of context length of the input.}
\label{fig:TTFT_comparison}
\end{figure}

Existing methods address the latency and memory bottlenecks of long-context inference at two main points in the pipeline. Some methods reduce inference cost within the model through KV cache reduction, token eviction, sparse attention, or semantic cache management \citep{zhang2023h2o, xiao2025duoattention, yan2025adamas}. These techniques modify attention computation or cache management during inference, reducing decoding-time latency and memory footprint. However, they require integration into the LLM inference stack, and the full prompt must still be processed during prefill before any internal reduction takes effect. Other methods instead compress the input prompt before inference \citep{xu2024recomp,hwang2025exit,liskavets2025cpc,zhang2026sentinel,jiang2023llmlingua,pan2024llmlingua2}. Because these methods shorten the text before it enters the target model, they can lower prefill computation, the KV cache size, and Time-To-First-Token (TTFT), while remaining complementary to KV cache methods that optimize later stages of inference. Figure~\ref{fig:TTFT_comparison} illustrates this pipeline-level distinction through a wall-time comparison of recent preprocessing and KV-cache methods at 32k and 128k context lengths.

Despite these advantages, preprocessing still leaves a distinct question of what structure the compressed prompt should adopt. Many extractive and pruning-based preprocessing methods operationalize this question by assigning scalar utility to local units, using retrievers, classifiers, contrastive encoders, proxy language models, attention signals, or query similarity, and then retaining high-utility units under a fixed budget \citep{xu2024recomp,hwang2025exit,liskavets2025cpc,zhang2026sentinel,jiang2023llmlingua,pan2024llmlingua2}. Even when the utility is context-aware, the final decision often remains a one-dimensional ranking over candidates, further discussed in Appendix~\ref{app:dimensionality}. This abstraction is efficient, but it leaves theme coverage implicit rather than allocating the budget across the recurring themes of the document. Classical retrieval and summarization work makes this coverage concern explicit by balancing relevance with novelty, diversity, or representativeness \citep{carbonell1998}. Under tight budgets, scalar ranking can therefore overrepresent the dominant theme and omit a rarer theme. In multi-hop QA \citep{yang2018hotpotqa, trivedi-etal-2022-musique}, for instance, it may retain several passages about the main entity while dropping the bridge sentence that connects it to the second entity. We refer to this coverage failure as \emph{theme collapse}.

To address this gap, we propose SALT, a model-agnostic extractive framework for preprocessing long-context inputs before inference. Rather than letting a single global ranking determine which sentences enter the compressed prompt, SALT first constructs a lightweight representation of the document's thematic structure and uses it to allocate the compression budget across recurring themes. Sentences are anchored by extracted keywords, and the frequency of these keywords across the document provides a document-derived salience signal rather than one learned from a task-specific ranker. The selected sentences are then reconstructed in document order as a plain-text prompt, making SALT usable with any downstream language model and complementary to architectural methods that target memory use or decoding latency. SALT is open-source and available on GitHub.\footnote{\url{https://github.com/oteomamo/SALT}}
Our main contributions are:
\begin{itemize}
    \setlength{\itemsep}{0.8pt}
    \setlength{\parskip}{0.5pt}
    \item We identify theme collapse, the loss of minor themes when scalar-ranked compressors are pushed to tight budgets, and reformulate extractive compression to allocate budget across document-derived lexical themes before sentence selection rather than as a post-hoc diversity correction.
    \item We propose SALT, a model-agnostic method for sentence-level prompt compression. SALT relies on a lightweight encoder to extract per-sentence keywords, organizes the salient ones into a trie ordered by sentence frequency, and allocates budget across trie branches before sentence selection.
    \item SALT outperforms prior preprocessing methods across accuracy, latency, and memory. It also matches state-of-the-art KV-cache techniques on latency and memory at a modest accuracy cost, while remaining usable across NVIDIA GPU generations.
\end{itemize}

\section{Related Work}
\label{sec:related}

Prior works have studied ways to reduce the cost of long-context inference through prompt reduction, cache compression, or attention-side optimization. We discuss these directions in this section.

Input-level prompt compression shortens the textual input before it reaches the target LLM. Sentence-level methods such as RECOMP, EXIT, CPC, and Sentinel select or rewrite textual units using learned retrievers and proxy-model signals~\citep{xu2024recomp,hwang2025exit,liskavets2025cpc,zhang2026sentinel}. Token-level compressors such as LLMLingua and LLMLingua-2 remove less informative tokens to achieve stronger compression ratios~\citep{jiang2023llmlingua,pan2024llmlingua2}. Under standardized evaluation, token-level pruning has also been found to trail extractive selection~\citep{jha2024characterizing}. These methods are closest to SALT in the inference pipeline because they reduce the number of tokens processed during prefill. However, their selection abstraction has limitations: most reduce compression to scoring independent units and filling a budget with high-scoring candidates. SALT instead builds a document-level lexical theme structure and allocates budget across theme branches before sentence selection.

Another line of work addresses long-context input inside the target model by compressing KV states and pruning attention or tokens to reduce memory or latency~\citep{zhang2023h2o,snapkv,xiao2025duoattention,fastkv,zhu2025sentencekv}. These approaches share SALT's efficiency goal but operate at a different stage of the inference stack. They require access to model internals or cache management, and the original prompt is still processed by the target model before or during internal compression. SALT is complementary: it outputs a shorter plain-text prompt before prefill, remains model-agnostic, and can be composed with cache-side methods.

SALT is also related to classical extractive summarization and diversity-aware retrieval. Maximal Marginal Relevance (MMR) balances query relevance against pairwise novelty, while submodular summarization optimizes coverage and diversity under a budget~\citep{carbonell1998,submodular}. These objectives show that compression should not greedily select only the highest-scoring units under an independent per-unit score. SALT differs in where diversity enters the decision. MMR and submodular objectives typically add novelty or set coverage during candidate selection after units have been scored. SALT makes coverage the primary allocation objective: budget is first distributed over document-induced lexical theme branches, and sentences are selected within those branches afterward. Thus, SALT performs pre-prefill extractive compression whose selection unit is the sentence but whose allocation unit is the lexical theme branch. This distinction is important under tight compression budgets, where late-stage diversity penalties cannot recover branches that never receive budget in the first place.

\section{Method}
\label{sec:method}

SALT compresses a document into a sentence-level subset of bounded size while preserving coverage of its recurring lexical themes. 
We use lexical themes rather than semantic themes because compression decisions must remain lightweight, reusable across turns, and stable under small embedding perturbations. Lexical recurrence provides a sparse and interpretable approximation of document thematic structure while avoiding repeated pairwise semantic comparisons during traversal.

\begin{figure*}[t]
  \includegraphics[width=\linewidth]{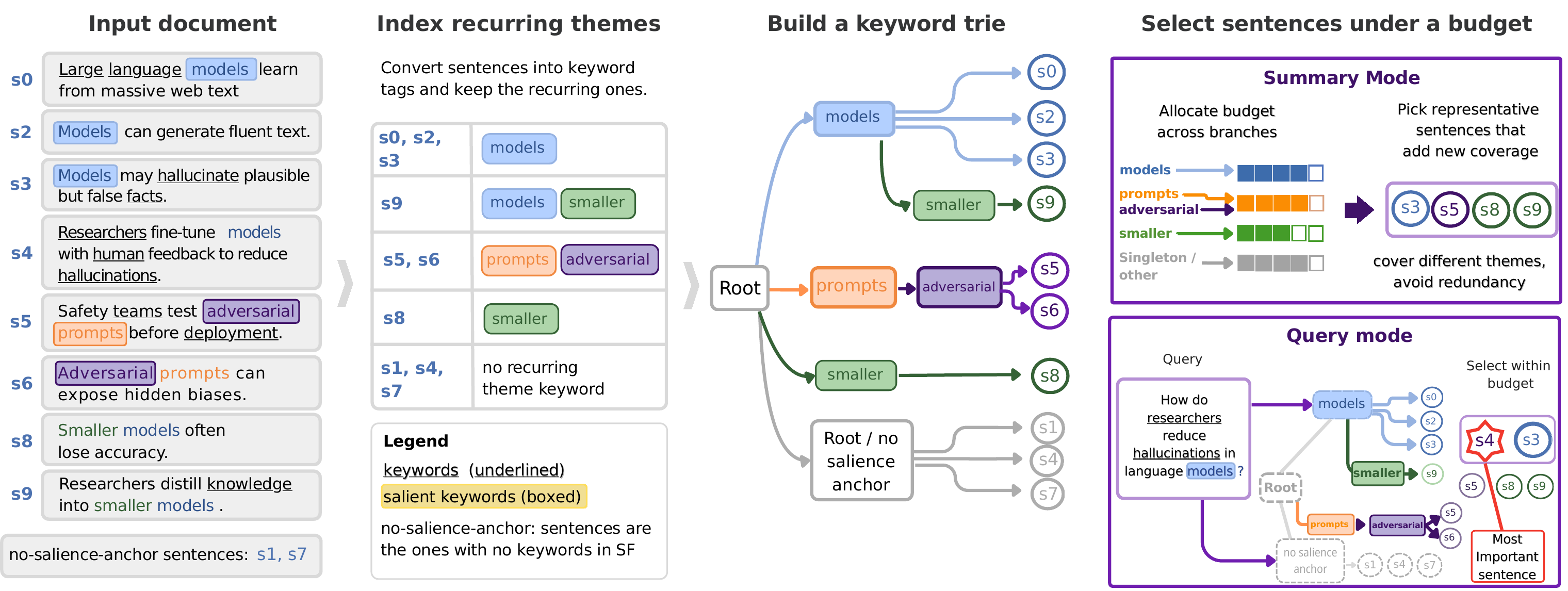}
\caption{Overview of the SALT pipeline.}
  \label{fig:method}
\end{figure*}

We organize SALT into two phases. An \emph{indexing} phase estimates these themes from sentence-level keyword statistics and organizes the document into a salience-aware lexical trie. A \emph{selection} phase chooses a sentence subset by traversing this structure under a target word budget, either unconditionally (\textit{summary} mode) or with a query-dependent bias (\textit{query} mode). These phases are separated for both logical and practical reasons. Logically, theme estimation and budget-constrained selection are distinct operations: the former defines the document structure, and the latter decides which parts of that structure can be represented under the budget. Practically, in multi-turn settings, the same indexed structure can serve successive queries without re-encoding the document. Designing the indexing artifact for repeated, query-conditioned access amortizes the encoding cost across turns and gives selection a uniform interface across modes. Figure~\ref{fig:method} gives the overview of SALT.

\subsection{Indexing}
\label{sec:method:indexing}
Given a document $D$ partitioned into $N$ sentences, we construct a candidate keyword ranking for each sentence using the lightweight open-source encoder BGE-small-en-v1.5 \citep{bge_embedding}. Specifically, we use [CLS] attention \citep{ding-luo-2021-attentionrank,devlin-etal-2019-bert} to rank content words within each sentence. This attention ranking is used only as a proposal signal, not as a faithful estimate of word importance; the number of words retained is determined by the reconstruction criterion described. Sentences are encoded within a 512-token window containing its surrounding context, while keyword selection is applied to the words of the target sentence. The full encoding protocol appears in Appendix~\ref{app:salience}.

Within each sentence, we rank content words by descending [CLS] attention and incrementally form prefixes of this ranking. At step $t$, we compute the cosine similarity $c_t$ between the mean embedding of the top-$t$ words and the mean embedding of the full sentence. Although $c_t$ tends to increase as more words are added, it is not guaranteed to be monotone: a newly added word can move the subset mean away from the full-sentence mean and produce a local dip. We therefore apply knee detection \citep{kneedle} to the monotone envelope
\begin{equation}
\bar{c}_t = \max_{\tau \le t} c_\tau, \qquad t = 1, \dots, n_i,
\label{eq:envelope}
\end{equation}
where $n_i$ is the number of content words in sentence $i$. The running maximum suppresses local dips and yields a non-decreasing reconstruction curve on which the kneedle criterion is 
defined. The resulting knee $k_i$ determines how many words are retained, but we cap it at $40\%$ of $n_i$ to prevent diffuse sentences from promoting most of their words into the keyword set. The top-$k_i$ content words in the attention ranking form $K_i$; all lower-ranked words are excluded from the keyword index.

Over the per-sentence keyword sets $\{K_i\}_{i=1}^N$, we compute the sentence frequency (SF) as
\begin{equation}
\label{eq:sf}
\mathrm{SF}(w) = \bigl|\{i : w \in K_i\}\bigr|,
\end{equation}
which measures how often a keyword appears as a selected anchor across sentences. A high sentence frequency indicates that the keyword participates in the document's recurring lexical structure, rather than appearing only in a local context. The salience set $\mathcal{S}$ retains keywords whose SFs are above the $p$-th quantile of this distribution (default $p=0.9$), as shown in Figure~\ref{fig:method}. This pruning removes low-frequency anchors and bounds the size of the trie, but it does not determine which themes are ultimately represented; coverage is imposed later by budget allocation across trie branches. Importantly, pruning affects only the indexing and scoring vocabulary. Since selection operates on whole sentences, any non-indexed word 
in a selected sentence remains in the compressed output. For selection, each retained keyword is assigned the normalized salience weight
\[
\widehat{\mathrm{SF}}(w) = \frac{\mathrm{SF}(w)}{\mathrm{SF}_{\max}},
\]
where $\mathrm{SF}_{\max}=\max_{u\in\mathcal{S}}\mathrm{SF}(u)$.

\subsection{The Keyword Trie}
\label{subsec:keyword_trie}

The salience set $\mathcal{S}$ and the per-sentence keyword sets define a reusable lexical representation of the document. We organize this representation as a \emph{keyword trie} $\mathcal{T}$ whose internal nodes are labeled by salience-set keywords and whose leaves store sentence identifiers. For each sentence $s_i$, we form $T_i = K_i \cap \mathcal{S}$, sort $T_i$ by decreasing sentence frequency, and insert the resulting sequence as a root-to-leaf path ending at $s_i$. Sentences with common highest-salience anchors will share a trie prefix, while branch points record where their secondary anchors diverge. This structure preserves keyword co-occurrences that would be lost by assigning each sentence only to its highest-frequency keyword.

Coverage is measured over keyword mass in a subtree rather than over the number of sentences selected from it. For a trie node $v$, let $\mathcal{D}(v)$ denote its descendant sentences and let
\[
\Gamma(v)=\bigcup_{s_i\in\mathcal{D}(v)}T_i
\]
be the keyword signature of its subtree. Given a partial output $\mathcal{R}$, the keywords covered at $v$ are
\[
C_v(\mathcal{R})=
\Gamma(v)\cap
\bigcup_{s_i\in\mathcal{R}\cap\mathcal{D}(v)}T_i .
\]
The uncovered mass of $v$ is then
\begin{equation}
U_v(\mathcal{R}) =
\sum_{w\in\Gamma(v)\setminus C_v(\mathcal{R})}
\widehat{\mathrm{SF}}(w).
\label{eq:uncovered_mass}
\end{equation}
Thus, a branch can be represented by a small number of sentences when those sentences cover its high-salience keyword signature. Additional sentences that repeat already covered anchors contribute little new mass, while sentences containing uncovered anchors reduce $U_v(\mathcal{R})$.

The trie also supports query-conditioned access without restricting traversal to ordinary prefix descent. Because the same keyword may appear at different depths depending on which anchors outrank it in each sentence, a query keyword activates all trie nodes with the corresponding label. Selection then unions the descendant regions of the activated nodes and applies the same coverage principle within those regions. This \emph{multi-anchor activation} allows query mode to recover sentences where a query-relevant keyword appears as either a primary anchor or a secondary co-occurrence.

The SALT trie is reusable across budgets, modes, and turns. Since indexing is independent of any particular query, successive queries over the same document can traverse the trie while changing only the activated nodes and selection scores. This separation amortizes the encoding cost across repeated accesses and gives summary-mode and query-mode compression a common document representation.

\subsection{Budget-Constrained Selection}
\label{sec:method:selection}

Selection traverses the trie under a target word budget $B$ and returns a sentence subset $\mathcal{R}$ in original document order, where
$\mathcal{R}$ denotes the current partial output and is updated after each admitted sentence. For a  sentence $s_i$ considered inside branch $b$, SALT scores the sentence by the reduction it would produce in the branch's uncovered mass:
\begin{equation}
\Delta_b(s_i \mid \mathcal{R})
=
U_b(\mathcal{R}) -
U_b(\mathcal{R} \cup \{s_i\}).
\label{eq:marginal-gain}
\end{equation}
The sentence score is calculated as
\begin{equation}
\mathrm{score}_b(s_i \mid \mathcal{R})
=
\Delta_b(s_i \mid \mathcal{R}) L(n_i) I(s_i),
\label{eq:score}
\end{equation}
where $L(n_i)$ favors moderate-length sentences and $I(s_i)$ encodes sentence-level priors such as position and, in query mode, lexical match. Because the gain is marginal, sentences that repeat already covered anchors lose value, while sentences that cover new branch anchors remain competitive. Algorithm~\ref{algorithm:salt} gives the unified procedure.

\begin{algorithm}[t]
\DontPrintSemicolon
\SetKwInOut{Input}{Input}
\SetKwInOut{Output}{Output}
\caption{SALT selection.}
\label{algorithm:salt}
\Input{Trie $\mathcal{T}$, keyword sets $\{K_i\}$, budget $B$, optional query $x$.}
\Output{Sentence subset $\mathcal{R}$ with $\mathrm{cost}(\mathcal{R}) \le B$.}
\BlankLine
$\mathcal{R} \gets \emptyset$;\quad $\mathcal{S}_{\mathrm{eff}} \gets \mathcal{S}$\;
\If{$x$ is given}{
    $K_q,\mathbf{e}_q \gets \mathrm{QueryIndex}(x)$\;
    $\mathcal{S}_{\mathrm{eff}} \gets \mathcal{S} \cup (K_q \cap \bigcup_i K_i)$\;
    $\mathcal{A} \gets \mathrm{QueryAnchors}(\mathcal{T}, K_q, \{K_i\})$\;
    $\mathcal{R} \gets \mathrm{AnchorPhase}(\mathcal{A}, K_q, \mathbf{e}_q, \beta_q B)$\;
}
$B_\star \gets B-\mathrm{cost}(\mathcal{R})$\;
$\mathcal{B} \gets \mathrm{ActiveBranches}(\mathcal{T},\mathcal{R})$\;
$\{\beta_b\}_{b\in\mathcal{B}} \gets \mathrm{BranchAllocate}(\mathcal{B},B_\star)$\;
$\mathcal{R} \gets \mathrm{BranchPhase}(\mathcal{T},\mathcal{R},\{\beta_b\})$\;
$\mathcal{R} \gets \mathrm{GlobalFill}(\mathcal{T},\mathcal{R},B)$\;
\Return $\mathcal{R}$ sorted in document order\;
\end{algorithm}

In summary mode, $\mathcal{S}_{\mathrm{eff}}=\mathcal{S}$, selection starts from the root. SALT first allocates the residual budget $B_\star$ across active depth-$1$ branches, then selects sentences within each branch using Eq.~\ref{eq:score}. The branch quota is a fixed floor plus a residual share proportional to $(M_b+\epsilon)^\alpha$, where $M_b$ is the branch's remaining uncovered mass and $0<\alpha<1$. The floor reserves capacity for low-mass branches, while the sublinear exponent compresses high-mass branches. This allocation counters theme collapse by reserving coverage across recurring themes before local sentence scores can commit the budget to a dominant branch. Any unused budget is assigned by a final \textit{GlobalFill} pass (Algorithm~\ref{algorithm:salt}).

In query mode, SALT extracts query keywords $K_q$ and the embedding $\mathbf{e}_q$. Query keywords already in $\mathcal{S}$ activate matching trie nodes; query keywords pruned from $\mathcal{S}$ but present in some $K_i$ are reactivated through the stored sentence-keyword index. The effective salience set is therefore $\mathcal{S}_{\mathrm{eff}}=\mathcal{S}\cup(K_q\cap\bigcup_i K_i)$, and query keywords receive larger mass when computing $U_b$.  The anchor phase ranks activated candidates based on lexical-anchor overlap with $K_q$ and positive embedding similarity to $\mathbf{e}_q$, admitting the top candidates and their immediate neighbors up to $\beta_q B$. The remaining budget is allocated using BranchAllocate and GlobalFill procedures as summary mode. As a result, $\beta_q$ controls the relevance--coverage trade-off: smaller values favor summary-style compression, while larger values prioritize query-aligned evidence.

\begin{table*}[th]
\centering
\small
\caption{LongBench results on LLaMA-3.1-8B-Instruct (20\% retention budget).}
\label{tab:longbench_20_llama}
\begin{tabular}{l *{7}{c}}
\toprule
\textbf{Method}
& \textbf{Single-Doc QA}
& \textbf{Multi-Doc QA}
& \textbf{Summarization}
& \textbf{Few-Shot}
& \textbf{Synthetic}
& \textbf{Code}
& \textbf{Avg.} \\
\midrule
\multicolumn{8}{c}{\textbf{Llama-3.1-8B-Instruct}} \\
\midrule
Full-context
& 43.58 & 44.65 & 29.22 & 69.48 & 54.21 & 60.01 & 50.19 \\
\midrule
\multicolumn{8}{c}{\textbf{KV Cache Methods (20\%)}} \\
\midrule
SnapKV        & 43.29 & 43.80 & 26.69 & 67.96 & 53.46 & 51.91 & 47.25 \\
FastKV        & 43.33 & 44.37 & 26.73 & 68.70 & 53.20 & 52.46 & 47.54 \\
SentenceKV    & 39.24 & 43.82 & 28.18 & 69.26 & 53.24 & 47.33 & 46.42 \\
DuoAttention  & 34.71 & 36.67 & 24.30 & 58.02 & 50.81 & 54.33 & 41.96 \\
\midrule
\multicolumn{8}{c}{\textbf{Preprocessing Methods (20\%)}} \\
\midrule
LLMLingua-2   & 30.28 & 30.39 & 24.04 & 19.26 & 12.62 & 39.57 & 26.02 \\
EXIT          & 31.30 & 23.37 & 25.14 & 58.56 & 13.50 & 37.34 & 32.30 \\
RECOMP        & 35.88 & 41.09 & 25.47 & 52.06 & 50.77 & 37.36 & 39.98 \\
CPC           & 38.91 & 39.42 & 24.97 & 49.11 & 51.50 & 21.84 & 37.74 \\
Sentinel      & 39.85 & 41.66 & 26.15 & 38.78 & 51.55 & 38.83 & 39.47 \\
\rowcolor{blue!10}
SALT          & 40.05 & 41.42 & 26.95 & 62.21 & 53.37 & 37.06 & 43.51 \\
\bottomrule
\end{tabular}

\end{table*}

\section{Experimental Evaluation}
\label{sec:evaluation}

\paragraph{Datasets and Models}
We evaluate accuracy on the English subset of LongBench \citep{bai2024longbench}, covering six task categories: Single-Doc QA, Multi-Doc QA, Summarization, Few-Shot Learning, Synthetic, and Code, as well as QuALITY \citep{pang2022quality} and RULER \citep{hsieh2024ruler} to assess long-context reasoning and retrieval. All methods use identical prompts and evaluation metrics following lm-eval-harness \citep{eval-harness}, ensuring differences are attributable solely to the compression method. All evaluated datasets include query except for the summarization section of the LongBench. For latency and memory profiling, we construct long-context inputs from the PG19 dataset \citep{pg19}. We present all main results on Llama-3.1-8B-Instruct \citep{dubey2024llama3herdmodels}, with additional results on Ministral-8B-Instruct-2410 \citep{ministral8b2024} and full percentage sweep included in the Appendix.

\paragraph{Hardware} We conduct most experiments on a compute node equipped with an NVIDIA H100 GPU and AMD EPYC 7R13 processor, except those in Section \ref{sec:eval-hardware-portability}, where we additionally benchmark on NVIDIA V100, A100, and B200 GPUs to assess performance across hardware generations. 

\paragraph{Configurations}
For comparative methods, we follow their published configurations with minimal changes for fair comparison. 
SnapKV runs at its reported defaults under a 20\% budget. FastKV uses a 60\% prefill and 20\% decode budget, matching its paper. DuoAttention is run at 20\% rather than its paper's best 50\% setting to keep methods budget-matched. SentenceKV uses the original implementation unchanged. All KV-cache methods use FlashAttention2 \citep{dao2024flashattention2}. EXIT splits documents into sentences with spaCy \citep{honnibal2020spacy} and scores each with the LoRA-tuned Gemma-2B classifier at threshold 0.5. RECOMP is used unchanged. CPC uses the released pretrained LoRA with a local Llama-3.1-8B-Instruct answer generator. Sentinel matches the paper's Qwen-2.5-0.5B proxy and trained detector. Minor preprocessing adjustments were made to Sentinel to prevent OOM on long inputs; details are provided in Appendix~\ref{app:accuracy}.
\subsection{Accuracy}\label{sec:eval:accuracy}
We report LongBench accuracy at a 20\% memory budget across all methods, the regime where compression is most aggressive and methodological differences are most informative. The current state-of-the-art KV-cache methods SnapKV and FastKV recover within roughly two to three points of the full-context baseline, while SALT provides similar accuracy for most subcategories, with code degrading the overall score the most. The gap is structural. KV-cache methods prune after the model has read the full prompt and use the model's own attention as a salience signal at token granularity, whereas preprocessing methods must commit up front using an external relevance estimate. Within the preprocessing methods, SALT leads on average and wins across most categories for the 20\% budget, showing it preserves the context and important sections of the input that causes other preprocessing methods to break. All preprocessing methods underperform on Code, where several KV-cache methods match full context, because code is line and token sensitive while preprocessing methods operate at sentence or paragraph level. Notably, SentenceKV, the one KV-cache method aggregating at sentence granularity, shows the same Code weakness, supporting a granularity explanation rather than a preprocessing versus KV-cache one. More results in Appendix~\ref{app:accuracy}. 

To directly test the coverage mechanism, we measure retained sentence-frequency (SF) keyword mass across budgets and compare SALT with a scalar-selection variant. SALT exceeds 90\% SF-mass coverage by 20\% retention on GovReport and QMSum, while GovReport shows the largest branch-allocation advantage at the tightest budget. A separate evaluation using KMeans clusters of sentence embeddings shows the same coverage advantage, and higher per-document SF coverage is associated with higher ROUGE-L. Full results and sensitivity analyses appear in Appendix~\ref{app:coverage-analysis}.

\subsection{End-to-End Efficiency at Scale}
\label{sec:eval:efficiency}

\begin{figure}[t]
  \centering
  \begin{subfigure}[t]{0.49\linewidth}
    \centering
    \includegraphics[width=\linewidth]{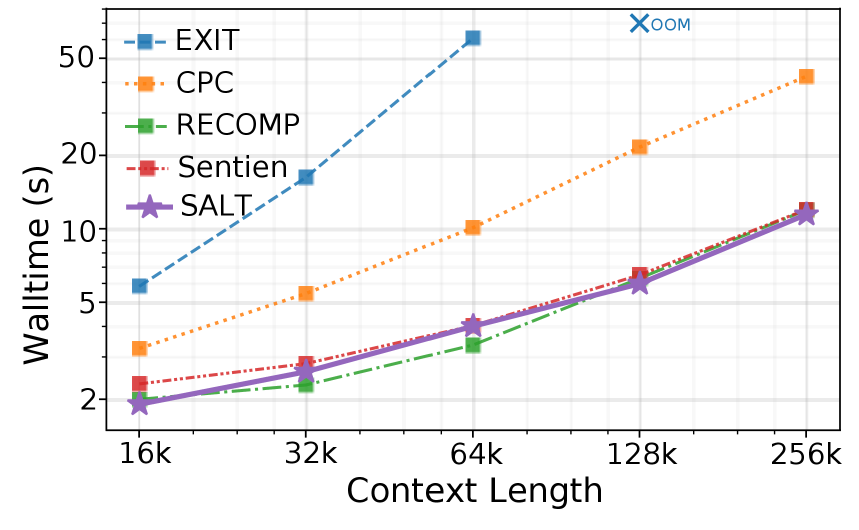}
    \caption{Preprocessing}
    \label{fig:eff-wt-prep}
  \end{subfigure}
  \hfill
  \begin{subfigure}[t]{0.49\linewidth}
    \centering
    \includegraphics[width=\linewidth]{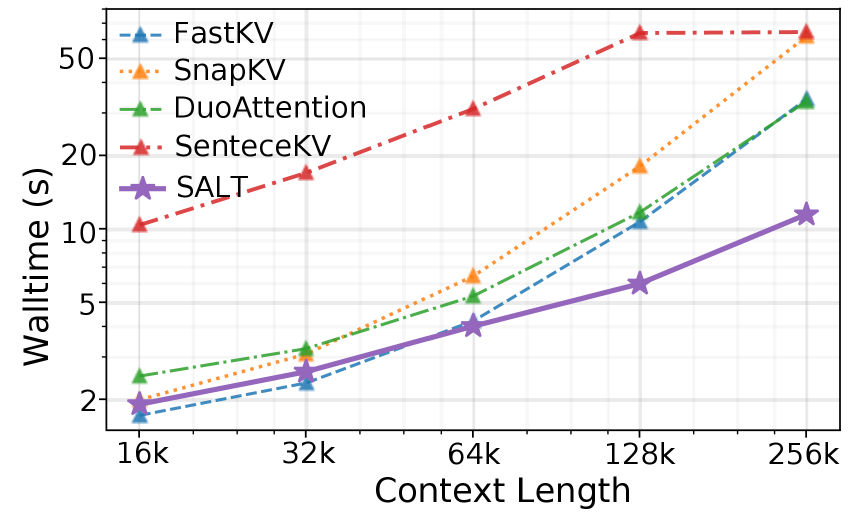}
    \caption{KV-cache}
    \label{fig:eff-wt-kv}
  \end{subfigure}

  \vspace{0.6em}

  \begin{subfigure}[t]{\linewidth}
    \centering
    \includegraphics[width=\linewidth]{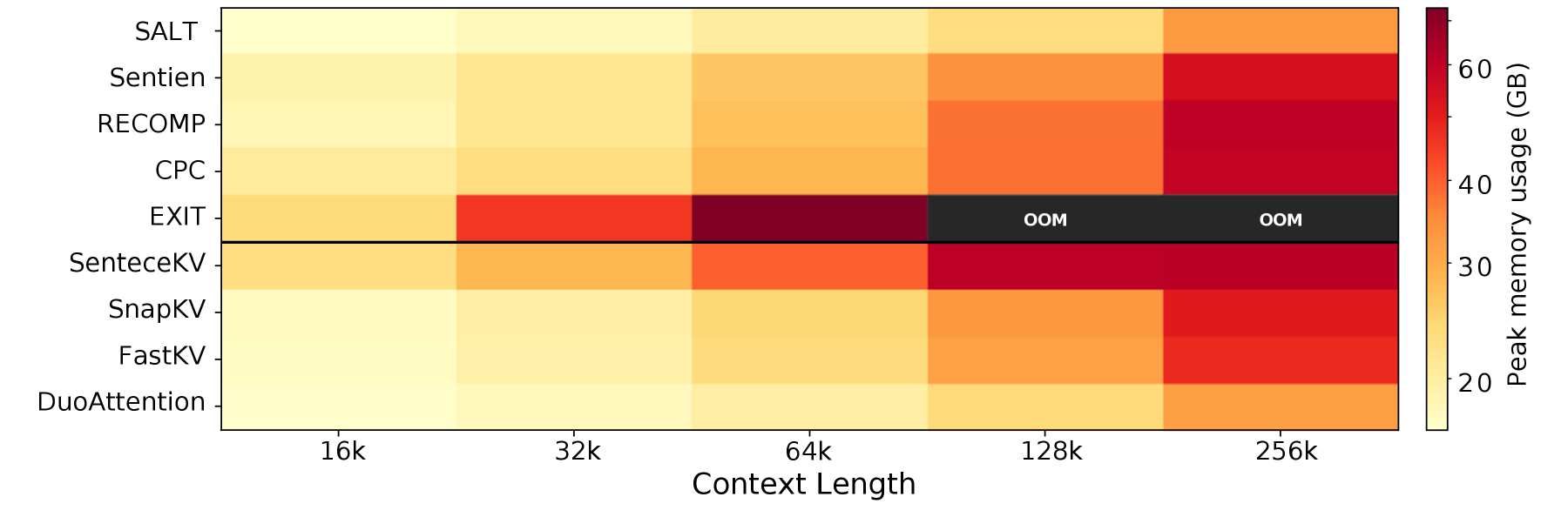}
    \caption{Peak GPU memory across methods and context lengths.}
    \label{fig:eff-mem}
  \end{subfigure}

  \caption{End-to-end efficiency of SALT against preprocessing and KV-cache baselines on Llama-3.1-8B across context lengths from 16k to 256k tokens. Memory and Walltime are in log scale.}
  \label{fig:efficiency-scale}
\end{figure}

Reducing the prompt only pays off if the reduction itself is cheap enough to run at scale. Preprocessing methods that score every chunk with an auxiliary model introduce overhead that grows with the input, while KV-cache methods inherit the cost of attention over the full prompt before they can compress it. We benchmarked SALT against four preprocessing baselines (EXIT, CPC, RECOMP, Sentinel) and four KV-cache baselines (FastKV, SnapKV, DuoAttention, SentenceKV) on Llama-3.1-8B at context lengths from 16k to 256k tokens, measuring peak GPU memory and walltime that includes a 128-token decode per prompt so that the cost of generation is also captured. Detailed values are reported in Appendix~\ref{app:efficiency}.

Figures~\ref{fig:eff-wt-prep} and~\ref{fig:eff-wt-kv} show the latency picture. Among preprocessing methods, those that delegate selection to an auxiliary model (e.g.,\ EXIT, CPC) inherit that model's forward pass at every input and degrade quickly as context grows. KV-cache methods generally amortize their selection at prefill and add little additional cost during decode; SentenceKV is the exception, as it performs sentence-level selection at prefill and continues to act through decode, paying extra cost on every generated token. Figure~\ref{fig:eff-mem} shows a similar separation on memory, and across both axes SALT remains among the cheapest methods at every length tested.

\subsection{Hardware Portability Across Context Lengths}
\label{sec:eval-hardware-portability}

\begin{figure}[t]
  \centering
  \includegraphics[width=\linewidth]{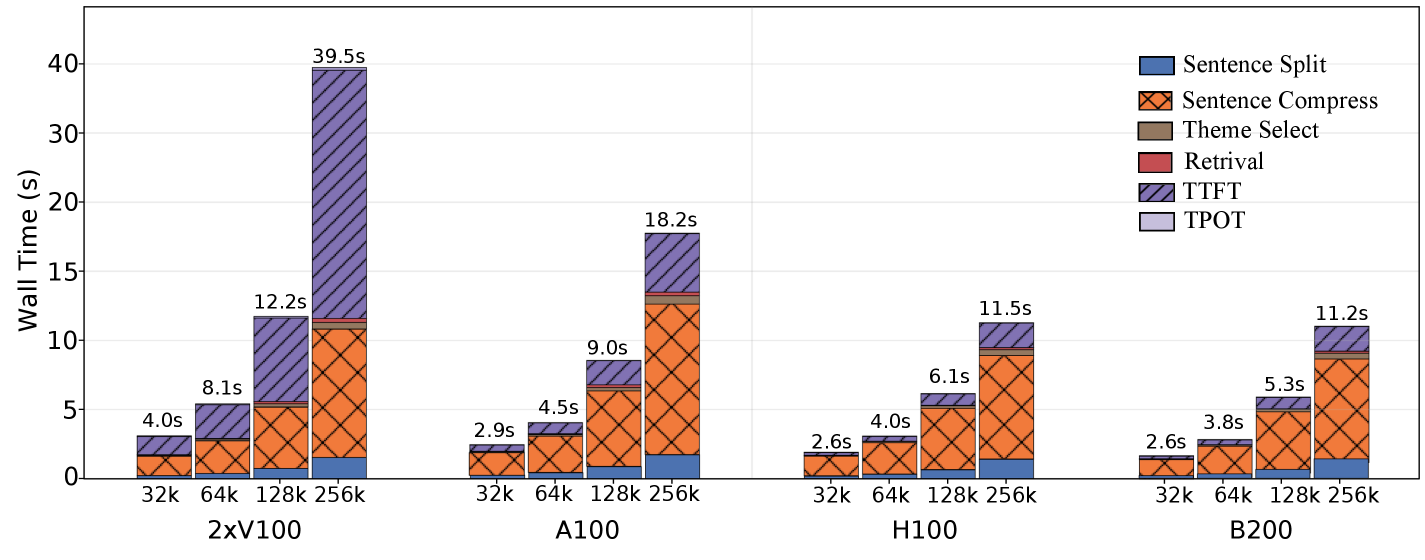}
  \caption{End-to-end latency and TPOT of SALT on Llama-3.1-8B-Instruct across different NVIDIA GPUs and raw-context lengths at a 20\% retention budget.}
  \label{fig:hardware-portability-latency}
\end{figure}

\begin{table}[t]
\centering

\small
\setlength{\tabcolsep}{6pt}
\caption{Peak GPU memory (GB) of SALT across prompt lengths and NVIDIA GPU architectures.}
\label{tab:memory-main}
\begin{tabular}{lcccc@{\hskip 1.2em}c}
\toprule
 & \multicolumn{4}{c}{Context length (tokens)} & Scaling \\
\cmidrule(lr){2-5} \cmidrule(lr){6-6}
GPU       & 32k   & 64k   & 128k  & 256k  & 256k / 32k \\
\midrule
V100x2    & 20.10 & 22.50 & 27.87 & 40.22 & 2.00$\times$ \\
A100     & 18.30 & 20.65 & 24.14& 33.29 & 1.81$\times$ \\
H100     & 17.90 & 20.20 & 23.30 & 32.80 & 1.83$\times$ \\
B200     & 16.41 & 17.86 & 20.67 & 25.72 & 1.57$\times$ \\
\bottomrule

\end{tabular}
\end{table}
Most reduction methods depend heavily on newer GPU architectures to manage memory during the prefill phase. Systems like SnapKV fundamentally require custom Triton kernels and FlashAttention compatibility, strictly mandating modern accelerators to execute efficiently. Conversely, input level preprocessing methods bypass these limits but often introduce new dependencies by relying on external proxy models to run their selection algorithms.

To demonstrate how our approach avoids these bottlenecks, we evaluated SALT using an 8B backbone at a fixed 20\% retention budget on PG19 datasets with contexts scaling up to 256k tokens. We compared four distinct GPU generations: a legacy Volta setup using two V100 units (32GB each) from 2017 against modern Ampere A100, Hopper H100, and Blackwell B200 accelerators.

Figure \ref{fig:hardware-portability-latency} shows that the older Volta hardware successfully processes 256k tokens. SALT is highly adaptable and does not depend on specific accelerator libraries. Furthermore, because SALT outputs a standard plain text prompt prior to prefill, it is not inherently restricted to NVIDIA GPUs, though we currently lack access to alternative architectures to present data on them. Detailed normalized walltime data is provided in Appendix \ref{app:accuracy}.

\subsection{Per-Turn Cost in Extended Interactions}
\label{sec:eval:conversational}

Single-turn benchmarks understate the compute footprint of compression in realistic deployment, where the same document is queried repeatedly across a conversation. We evaluate this regime on QuALITY with a 50-article subset and 972 question turns at a 20\% budget, comparing an uncompressed baseline, FastKV, RECOMP, and SALT on Llama-3.1-8B-Instruct. The full setup and per-turn timing breakdown are in Appendix~\ref{app:efficiency}. Figure~\ref{fig:multiturn-quality} shows that per-turn accuracy is essentially flat across 19 turns for every method, with SALT matching the baseline and FastKV throughout.
\begin{figure}[t]
  \centering
  \begin{subfigure}[t]{0.48\columnwidth}
    \centering
    \includegraphics[width=\linewidth]{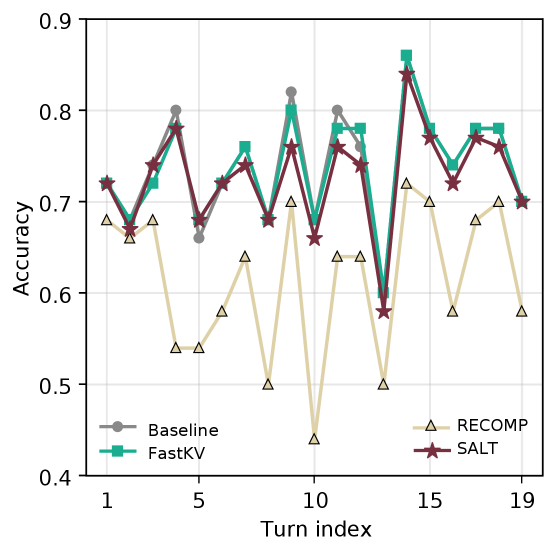}
    \caption{Accuracy}
    \label{fig:multiturn-acc}
  \end{subfigure}\hfill
  \begin{subfigure}[t]{0.48\columnwidth}
    \centering
    \includegraphics[width=\linewidth]{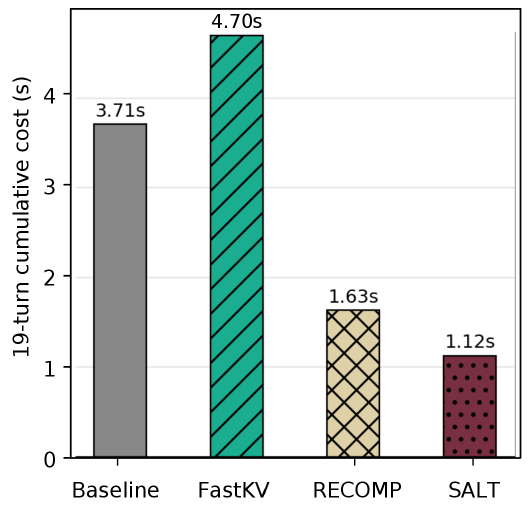}
    \caption{Cost}
    \label{fig:multiturn-cost}
  \end{subfigure}
  \caption{QuALITY at 20\% budget, 50 articles, 972 turns. (a) Per-turn accuracy is bounded for all methods. (b) Cumulative compute over a 19-turn conversation diverges sharply.}
  \label{fig:multiturn-quality}
\end{figure}

\subsection{Needle-in-a-Haystack (NIAH)}
\label{sec:eval:quality}

\begin{figure}[t]
    \centering
    \begin{minipage}{0.50\linewidth}
        \centering
        \includegraphics[width=\linewidth]{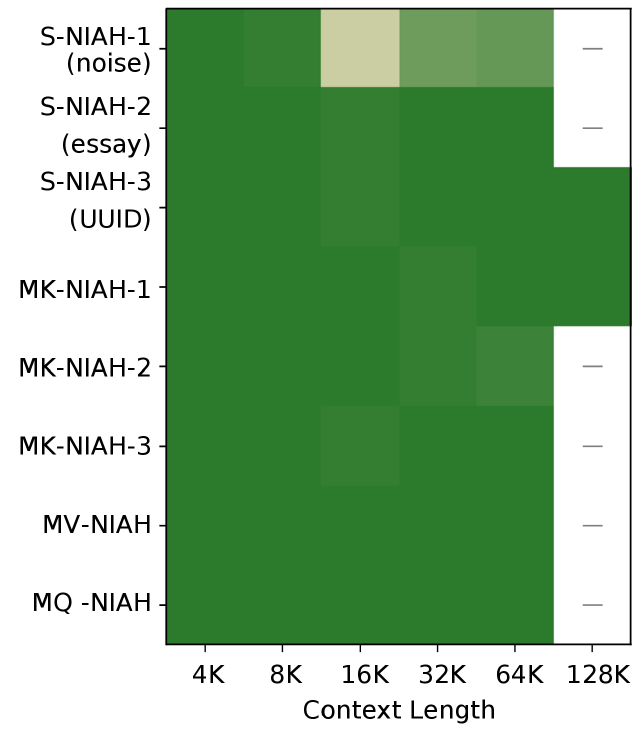}
        \subcaption{Baseline}
        \label{fig:ruler-heatmap-uncompressed}
    \end{minipage}
    \hfill
    \begin{minipage}{0.48\linewidth}
        \centering
        \includegraphics[width=\linewidth]{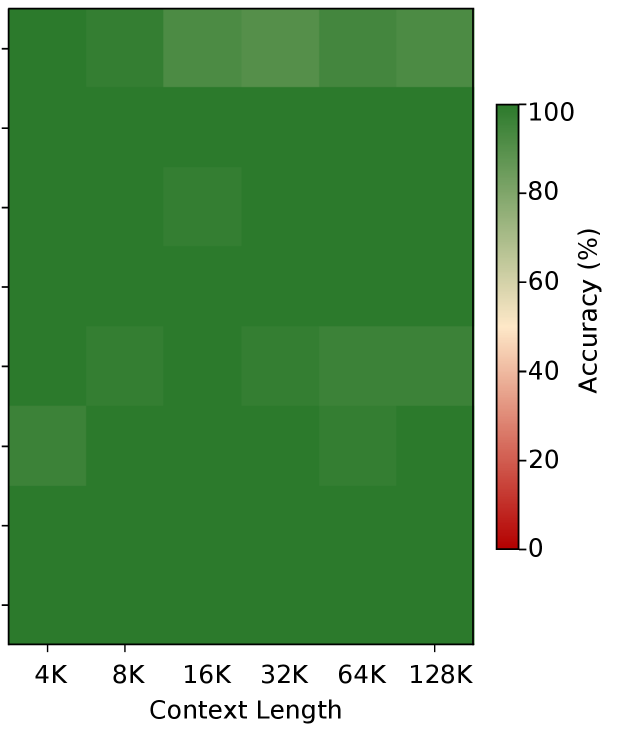}
        \subcaption{SALT}
        \label{fig:ruler-heatmap-salt}
    \end{minipage}
    \caption{%
Needle-in-a-Haystack (NIAH) results from RULER on Llama-3.1-8B-Instruct. The baseline uses the uncompressed prompt, while SALT retains 20\% of the source context before inference. White cells indicate unavailable full-context runs whose final tokenized prompts exceed the model context window.}
    \label{fig:ruler-heatmap-comparison}
\end{figure}

We evaluate SALT on the NIAH retrieval benchmark from RULER \citep{hsieh2024ruler}, which embeds a specific fact ("needle") within a long distractor context and queries the model to retrieve it. We report the eight NIAH variants spanning single-needle, multi-key, multi-value, and multi-query retrieval. As shown in Figure \ref{fig:ruler-heatmap-comparison}, SALT preserves NIAH accuracy across all context lengths matching the uncompressed baseline on NIAH tasks. SALT matches the uncompressed baseline at every length where the full-context prompt fits the model window. At the longest settings the baseline is unavailable (white cells, Figure \ref{fig:ruler-heatmap-uncompressed}) because the tokenized prompt exceeds the window while SALT's compressed prompt still fits. We report those cells without a baseline comparison.

\section{Conclusion}
We introduced SALT, a lightweight, model-agnostic extractive framework that reframes prompt compression as preserving thematic coverage under a fixed budget, using sentence frequency of lexical keywords as a reusable proxy. By organizing per-sentence keywords into an SF-ordered trie, SALT allocates budget across recurring themes before sentence selection and mitigates the theme collapse caused by scalar ranking. The trie supports both summary-mode traversal and multi-anchor query retrieval, enabling multi-turn dialogue without re-encoding the document, while outputting plain text compatible with any downstream LLM and complementary to KV-cache optimizations.

\section*{Limitations}

SALT has limitations on code-heavy tasks, where it underperforms both the full-context baseline and KV-cache methods that retain token-level granularity. Because SALT operates at the sentence level, it cannot reliably distinguish where one code unit ends and the next begins, often grouping or splitting code fragments in ways that break their semantics during retrieval. Lexical keyword extraction is also a poor fit for source code, where identifiers, operators, and structural tokens carry meaning that the [CLS] attention signal was never trained to capture. This is not unique to SALT as all preprocessing methods in our comparison degrade on Code for the same granularity and tokenization reasons.

SALT is also limited by its sentence-level granularity. Across our evaluations, the amount of task-relevant context varies substantially across datasets and often occupies only a small fraction of the full input. Query-based tasks may depend on only a few sentences, whereas summarization can require broader coverage, with useful retained spans often falling within the 5--20\% range. We use 20\% as a common evaluation budget rather than as a lower bound on SALT's compression. At more aggressive budgets, whole-sentence selection provides less flexibility than token-level KV-cache methods. SALT trades this finer granularity for a model-agnostic, pre-prefill compression path that remains composable with KV-cache methods downstream.

Finally, our evaluation is limited to English. All benchmarks (the English subset of LongBench, QuALITY, RULER, and PG19) are English-language, and our keyword extractor, BGE-small-en-v1.5, is an English-only encoder.

\section*{Acknowledgments}

This work was supported in part by the U.S. National Science Foundation
award 2403089. The research was also supported by the FSU-AWS award and
used the NoleLand facility funded by the U.S. National Science Foundation
grant CNS-1822737.

Following the ACL policy on AI assistance, we used AI assistants in a limited capacity during the preparation of this work.
Their use was restricted to editorial assistance, including grammar correction,
rephrasing for clarity, and \LaTeX{} formatting, as well as auxiliary coding
support for evaluation scripts, plotting routines, and debugging. The research
contributions in this paper, including the method design, experimental protocol,
analysis, and interpretation of results, are entirely our own. We did not use
AI assistants to generate scientific claims, develop the core method, or produce
experimental results. We reviewed and edited all AI-assisted content and take
full responsibility for the final manuscript.

\bibliography{references}

\clearpage
\appendix

\section{Dimensionality of Scalar Preprocessing}
\label{app:dimensionality}

We now clarify the sense in which scalar preprocessing turns compression into a one-dimensional selection problem. The claim is not that existing compressors use simple models or ignore context. Many preprocessing methods use contextual encoders, retrievers, proxy language models, or query-aware signals. The limitation appears at the selection interface. Once each sentence, passage, or token is assigned a single utility value, the budget is usually filled by comparing candidates along that value, sometimes with a later diversity correction. This reduces a document with several recurring themes to a flat list of candidates.

A scalar list is efficient, but it does not specify how the budget should be distributed across the themes of the document. If many high-scoring candidates come from the dominant theme, top-ranked selection can spend most of the budget on that theme. Lower-frequency themes can then disappear, even when they are needed to represent the document or support downstream reasoning. We refer to this coverage failure as theme collapse.

Classical work in diversity-aware retrieval and summarization, including embedding-based selection over centroids or nearest neighbors, has long argued that relevance alone is not sufficient under a budget and that novelty, diversity, or coverage must enter the objective \citep{carbonell1998, submodular}. SALT adopts this concern as an input-level compression principle. Instead of relying on a final ranked list, whether from a scalar utility model or from geometric proximity to a document centroid, to preserve coverage implicitly, SALT allocates budget across document-induced theme branches before selecting sentences within those branches.

To test whether scalar order alone is a reliable guide, we run a diagnostic summarization ablation at a 20\% token budget. All compressed variants use the same embedding model, grouping procedure, and document-order reconstruction. The difference is how candidates are selected from the score distribution. As shown in Table~\ref{tab:dimensionality_compact}, selecting only the top-ranked centroid candidates is not consistently superior. These results do not imply that every error comes from theme collapse. They show the narrower point needed here. A single global score is not a complete description of a candidate's value under a tight budget, because useful information can lie outside the top of the scalar ranking.
\begin{table}[t]
\centering
\caption{Diagnostic ROUGE-L results at a 20\% token budget. Full context is included only as an upper reference. All methods use document-order reconstruction.}
\label{tab:dimensionality_compact}
\small
\setlength{\tabcolsep}{6pt}
\begin{tabular}{l *{5}{c}}
\toprule
\textbf{Method} & \textbf{Gov.} & \textbf{QMS.} & \textbf{MNews} & \textbf{Avg.} \\
\midrule
Full context        & 35.14 & 25.78 & 26.73 & 29.22 \\
Random              & 29.63 & 21.73 & 22.02 & 24.46 \\
\midrule
kNN                 & 30.23 & 20.95 & 22.50 & 24.56 \\
Centroid            & 29.64 & 20.45 & 21.62 & 23.90 \\
Centroid + filt.    & 29.69 & 19.94 & 22.61 & 24.08 \\
Lead + centroid     & 29.23 & 20.73 & 22.87 & 24.28 \\
Spread selection    & 30.34 & 20.62 & 22.17 & 24.38 \\
\rowcolor{blue!10}
SALT (Trie)         & 32.77 & 23.99 & 24.08 & 26.95 \\
\bottomrule
\end{tabular}
\end{table}
This motivates explicit allocation across themes. SALT still uses sentence-level scores, but it does not let a single global ranking determine the compressed prompt. The trie first exposes recurring lexical branches in the document, and the retrieval procedure assigns budget across those branches. Sentence scores are then used within each branch. This separates the coverage decision from the local selection decision, reducing the chance that the dominant theme consumes the budget before minor themes can appear.

\section{Theme Coverage Analysis}
\label{app:coverage-analysis}

We directly evaluate whether SALT preserves recurring themes as the compression budget changes. We measure SF-mass coverage as $1-U_r(\mathcal{R})/U_r(\emptyset)$, where $r$ is the trie root and $U$ is defined in Eq.~\ref{eq:uncovered_mass}. As shown in Table~\ref{tab:theme_coverage}, SALT retains substantial thematic mass even at tight budgets. GovReport and QMSum both exceed 90\% coverage at 20\% retention, with smaller gains thereafter.

\begin{table}[t]
\centering
\caption{SF-mass coverage (\%) across compression budgets.}
\label{tab:theme_coverage}
\small
\setlength{\tabcolsep}{5pt}
\begin{tabular}{lrrrr}
\toprule
\textbf{Dataset} & \textbf{10\%} & \textbf{20\%} &
\textbf{30\%} & \textbf{40\%} \\
\midrule
GovReport & 81.6 & 91.4 & 95.9 & 97.9 \\
QMSum     & 80.9 & 92.7 & 97.0 & 98.7 \\
MultiNews & 70.6 & 82.5 & 88.8 & 93.0 \\
\bottomrule
\end{tabular}
\end{table}

We isolate the effect of branch allocation on GovReport by removing the depth-1 branch partition while keeping the remaining selection procedure fixed. Table~\ref{tab:coverage_ablation} shows that SALT consistently retains more SF mass than the scalar variant. The difference is largest at 10\% retention and decreases from 2.69 to 1.07 percentage points as the budget increases. This is consistent with branch allocation being most useful when multiple themes compete for a limited budget.

\begin{table}[t]
\centering
\caption{Branch-allocation ablation on GovReport. Coverage values are
SF-mass coverage; $\Delta$ denotes SALT minus the scalar variant.}
\label{tab:coverage_ablation}
\small
\setlength{\tabcolsep}{4pt}
\begin{tabular}{crrrr}
\toprule
\textbf{Budget} &
\textbf{SALT} &
\textbf{Scalar} &
\textbf{$\Delta$ Cov.} &
\textbf{$\Delta$ ROUGE-L} \\
\midrule
10\% & 0.8157 & 0.7888 & +2.69 pp & +0.23 \\
20\% & 0.9138 & 0.8966 & +1.72 pp & +0.04 \\
30\% & 0.9594 & 0.9449 & +1.45 pp & +0.37 \\
40\% & 0.9790 & 0.9683 & +1.07 pp & +0.00 \\
\bottomrule
\end{tabular}
\end{table}

Coverage also tracks downstream quality across documents. We measure this association using Spearman rank correlation, denoted by $\rho$. On GovReport, per-document SF-mass coverage is positively correlated with ROUGE-L at every budget, with $\rho=0.324$, $0.215$, $0.188$, and $0.192$ from 10--40\% retention ($p<0.05$ in all cases). Controlling for log document length and realized compression ratio gives partial correlations of $0.364$, $0.385$, $0.304$, and $0.276$, indicating that the association remains after controlling for these factors.

We also test whether the coverage effect extends beyond SALT's lexical theme definition. We cluster BGE \citep{bge_embedding} sentence embeddings with K-means using $k\in\{5,10,20\}$ clusters and count a semantic cluster as covered when at least one of its sentences is retained. These clusters are used only for evaluation and do not affect SALT's selection. As shown in Table~\ref{tab:semantic_coverage}, SALT covers more semantic clusters than the scalar variant across all budgets and clustering granularities.

\begin{table}[t]
\centering
\caption{SALT minus scalar semantic-cluster coverage on GovReport
(percentage points).}
\label{tab:semantic_coverage}
\small
\setlength{\tabcolsep}{5pt}
\begin{tabular}{lrrrr}
\toprule
 & \textbf{10\%} & \textbf{20\%} & \textbf{30\%} & \textbf{40\%} \\
\midrule
$k=5$  & +11.30 & +8.70  & +6.40  & +4.60 \\
$k=10$ & +13.30 & +9.45  & +8.05  & +6.30 \\
$k=20$ & +10.08 & +11.10 & +10.35 & +7.75 \\
\bottomrule
\end{tabular}
\end{table}

For $k=10$, the semantic-coverage difference is significant at every budget ($p<10^{-16}$). The effect also holds across all three clustering granularities, showing that the coverage advantage is not specific to SALT's SF-based theme definition. We additionally divide keywords within each document into SF terciles corresponding to minor-, mid-, and major-frequency themes. As the budget increases, SALT's advantage on major themes decreases from 2.03 percentage points at 10\% retention to 0.38 points at 40\%, while remaining positive for the mid- and minor-frequency bands. This is consistent with branch allocation shifting capacity toward themes that are more likely to be crowded out by a global scalar ranking.

\section{Keyword Extraction via Transformer Attention}
\label{app:salience}

We extract per-sentence keywords from the input document by repurposing the internal attention patterns of a pretrained transformer encoder. Rather than using the model's output embeddings in the standard way, we hook into two intermediate signals: (i)~the \texttt{[CLS]} token's attention weights from the final layer, which indicate per-token importance, and (ii)~the per-token hidden-state vectors, which provide contextual embeddings for measuring how well a keyword subset reconstructs the full sentence meaning. The extracted keywords and their attention-derived importance weights constitute the atomic representational units for all downstream processing.
 
\subsection{Model Selection}
 
We employ \texttt{BAAI/bge-small-en-v1.5} \citep{bge_embedding}, a 12-layer, 12-head BERT encoder with hidden dimension $d{=}384$ and approximately 33M parameters. This model was selected for three architectural and empirical reasons.
 
First, its compact architecture (33M parameters, 12 layers) provides a favorable trade-off between representational capacity and inference cost. With documents routinely exceeding 10{,}000 tokens, the model must process dozens of packed chunks per document. A larger encoder would increase latency without proportionate gains in attention quality for keyword extraction, which depends primarily on the final layer's attention distribution rather than deep semantic reasoning.
 
Second, BGE was trained via contrastive learning on sentence-level retrieval tasks, which directly optimizes the \texttt{[CLS]} token to aggregate discriminative sentence-level information. This training objective produces attention patterns where \texttt{[CLS]} selectively focuses on content-bearing tokens, precisely the signal we extract. Models trained with different objectives (e.g., masked language modeling alone) distribute \texttt{[CLS]} attention more uniformly, yielding less informative importance rankings.
 
Third, we evaluated BGE against GTE-small \citep{li2023towards} (identical BERT architecture, different training objective) on the same documents. Both models extract keywords via the same pipeline, but their attention distributions differ fundamentally. GTE's contrastive training with cosine similarity supervision produces more peaked attention, a small number of dominant topic terms receive disproportionate weight across most sentences. BGE's instruction-tuned retrieval training produces more distributed attention, assigning meaningful weight to a broader set of content words per sentence. In practice, GTE extracted ${\sim}20\%$ fewer unique keywords across the same document, with its top terms appearing at $2$--$3{\times}$ the frequency of BGE's. For keyword-based downstream processing, this concentration is disadvantageous: when most sentences share the same few high-weight keywords, it becomes difficult to distinguish which sentences cover which specific aspects of the document. BGE's broader vocabulary provides the granularity needed to differentiate sentence-level content.
 
\subsection{Dense Packing with Span-Local Renormalization}
 
A standard approach processes each sentence through the encoder independently. For a document of $N$ sentences averaging ${\sim}28$ tokens each, this requires $N$ forward passes, each utilizing under $6\%$ of the model's 512-token input capacity. Beyond the computational waste, isolated encoding deprives the model of cross-sentence context: the attention pattern for a sentence is computed without knowledge of what surrounds it, so the model cannot distinguish document-central terms from locally prominent but globally generic ones.
 
We address both limitations through \emph{dense packing}: consecutive sentences are greedily concatenated into 512-token chunks, with a 2-sentence overlap between adjacent chunks to ensure boundary sentences receive context from both directions. This reduces the number of forward passes by approximately an order of magnitude while exposing each sentence to its neighborhood during attention computation.
 
The key challenge is recovering per-sentence keyword rankings from a chunk-level attention distribution. When multiple sentences share a single \texttt{[CLS]} attention vector, tokens compete globally. A keyword in one sentence may receive low attention simply because a different sentence in the same chunk contains higher-salience terms. We resolve this through \emph{span-local renormalization}: for each sentence's token span $[s, e)$ within a chunk, the raw CLS attention scores are divided by their span sum:
\begin{equation}
\hat{a}_j^{(s_i)} = \frac{a_j}{\sum_{k=s}^{e-1} a_k}, \quad j \in [s, e)
\end{equation}
This produces a probability distribution that sums to~1 within each sentence. Critically, the global context that shaped the raw attention values is preserved, the model ``saw'' neighboring sentences when computing these scores, but the ranking is now relative to the sentence's own tokens. A term that the model considers important given the surrounding context will rank highly even if its raw score is modest compared to tokens in adjacent sentences.
 
For sentences appearing in two overlapping chunks, we retain the renormalized attention from the chunk where the sentence received the highest total raw attention mass (indicating the most informative context window), while raw attention scores are MAX-aggregated across chunks to preserve any importance signal observed in either context.
 
\subsection{Kneedle-Based Keyword Selection}
Given the per-word attention scores for a sentence, we determine how many words qualify as keywords. A fixed threshold or fixed percentage would ignore the natural variation in how attention distributes across sentences of different lengths and information densities. Instead, we use a data-driven cutoff based on the geometry of an accumulation curve.
 
Words are ranked by descending attention and incrementally added to a subset.
At step $t$, we compute the cosine similarity $c_t$ between the mean
hidden-state embedding of the accumulated subset and the mean embedding of the
full sentence. Although $c_t$ generally increases as more words are added, it
is not guaranteed to be monotone. We therefore construct the monotone envelope

\begin{equation}
\bar{c}_t = \max_{\tau \leq t} c_\tau,
\qquad t=1,\ldots,n_i,
\end{equation}

where $n_i$ is the number of content words in sentence $i$. The running maximum
suppresses local decreases while preserving the overall reconstruction trend.
We then apply the Kneedle algorithm~\citep{kneedle} to the normalized curve
$\{(t,\bar{c}_t)\}$ to identify the transition from rapid reconstruction gains
to diminishing returns. The resulting knee $k_i$ determines the number of
keywords retained, subject to the cap

\begin{equation}
k_i \leq \left\lfloor r_{\max} n_i \right\rfloor,
\qquad r_{\max}=0.4.
\end{equation}

The top-$k_i$ content words in the attention ranking form the sentence's
keyword set $K_i$, with each keyword retaining its associated attention weight
$a_k$.

\begin{figure}[t]
    \centering
    
    \begin{subfigure}{0.23\textwidth}
        \centering
        \includegraphics[width=\linewidth]{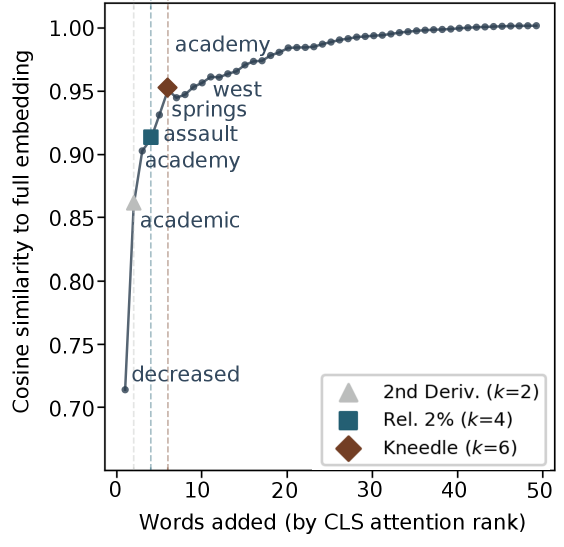}
        \caption{Cosine accumulation }
        \label{fig:cls-cosine}
    \end{subfigure}
    \begin{subfigure}{0.24\textwidth}
        \centering
        \includegraphics[width=\linewidth]{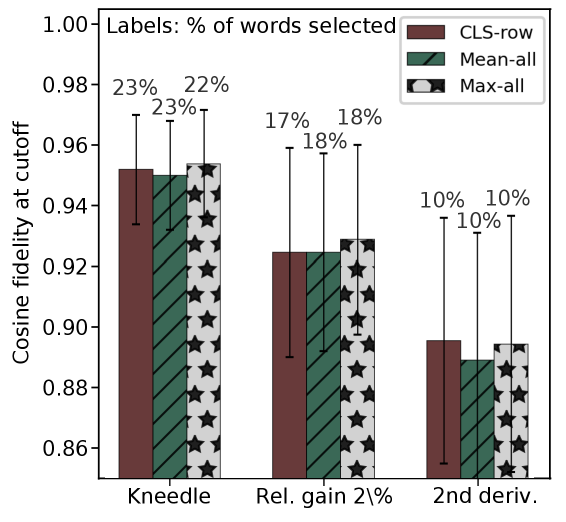}
        \caption{Attention source × cutoff}
        \label{fig:cls-attention}
    \end{subfigure}
    
    \caption{(a)~Cosine accumulation curve for a sample sentence: words added in CLS attention order progressively reconstruct the full-sentence embedding. The Kneedle cutoff (green) identifies the point of diminishing returns. (b)~Comparison of three attention sources (CLS-row, mean-all, max-all) and three cutoff methods. Labels indicate the fraction of words selected. CLS-row with Kneedle achieves high fidelity at low selection rate.}
    \label{fig:cls-combined}
\end{figure}

We evaluated two alternative cutoff methods. The \emph{relative gain drop} method (stop when the cosine gain at step $t$ falls below $2\%$ of the first step's gain) reacts to the initial steep drop and terminates prematurely, typically at $k{=}2$, selecting only ${\sim}7\%$ of words, which is insufficient to capture the sentence's informational breadth. The \emph{second-derivative} method (stop at the point of maximum deceleration) exhibits the same early-termination bias, since the largest deceleration in a concave curve always occurs at the transition from the first to second step. Kneedle, by contrast, detects the global shape transition rather than local rate changes, yielding a stable cutoff at ${\sim}21\%$ of words with cosine fidelity of ${\sim}0.95$ (Figure~\ref{fig:cls-combined}b).
 
We additionally compared CLS-row attention (our approach) against two alternative extraction signals: mean attention received by each token from all query positions, and maximum attention received from any single query position. CLS-row attention with Kneedle achieves the highest cosine fidelity at the lowest selection fraction, confirming that the \texttt{[CLS]} token's attention row is a more focused importance signal than aggregated all-token attention (Figure~\ref{fig:cls-combined}b).
 
\subsection{Output Format}
 
Phase~1 produces, for each sentence $s_i$, a keyword set $K_i$ with associated attention weights $\{a_k\}_{k\in K_i}$ and per-token hidden-state embeddings $\{\mathbf{h}_j\}$ from the final transformer layer. The keyword sets typically contain 5-7 content words per sentence (median~6 at average sentence length~28 words), with attention weights reflecting the model's assessment of each keyword's importance to the sentence-level representation.

\begin{table}[t]\centering\footnotesize
\setlength{\tabcolsep}{7pt}
\caption{Trie statistics averaged over 30 PG19 inputs per context length.
The trie is built once per document and is invariant to the compression
budget.}
\begin{tabular}{@{}lrrrr@{}}
\toprule
 & 32k & 64k & 128k & 256k \\
\midrule
Sentences & 1{,}783 & 3{,}365 & 7{,}738 & 15{,}411 \\
Salience set $|\mathcal{S}|$ & 176 & 388 & 770 & 1{,}164 \\
Trie nodes & 2{,}163 & 3{,}546 & 9{,}679 & 20{,}582 \\
Nodes per sentence & 1.21 & 1.05 & 1.25 & 1.34 \\
Depth-1 branches & 149 & 327 & 637 & 1{,}007 \\
Median path depth & 2.2 & 2.1 & 2.1 & 2.4 \\
Max path depth & 7.4 & 6.7 & 8.1 & 8.3 \\
No-anchor sent.\ (\%) & 8.7 & 10.6 & 8.7 & 7.6 \\
\bottomrule
\end{tabular}
\label{tab:trie-stats}
\end{table}

\section{Trie Size and Depth Across Context Lengths}
\label{app:trie-stats}

\begin{figure}[t]\centering
\begin{subfigure}{0.48\columnwidth}
  \includegraphics[width=\linewidth]{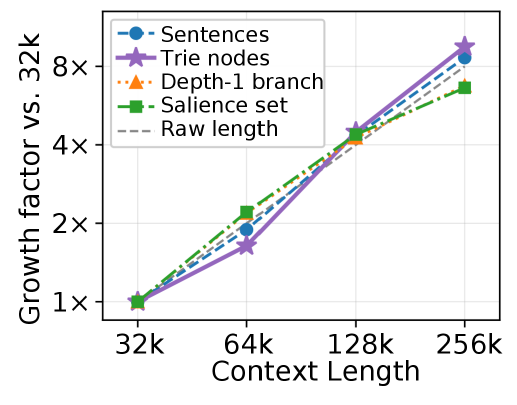}
  \caption{Growth vs.\ 32k}
\end{subfigure}\hfill
\begin{subfigure}{0.48\columnwidth}
  \includegraphics[width=\linewidth]{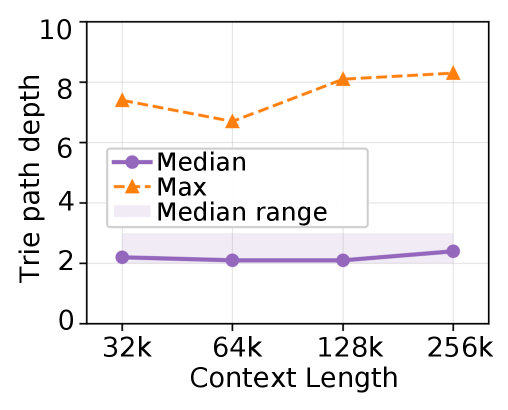}
  \caption{Path depth}
\end{subfigure}
\caption{Trie scaling on the PG19 profiling inputs. (a) Node count tracks
sentence count near-linearly, while depth-1 branches and the salience set
grow sublinearly relative to raw length (dashed). (b) Path depth is
invariant to context length (shaded: range of per-document medians).}
\label{fig:trie-scaling}
\end{figure}

To quantify how the index of Section~\ref{subsec:keyword_trie} scales, we build one trie per document over 30 PG19 inputs at each target length $L\in\{32\mathrm{k},64\mathrm{k},128\mathrm{k},256\mathrm{k}\}$ tokens. For each $L$, the inputs are formed as disjoint $L$-token windows cut from the train split in stream order, using the default indexing configuration of Section~\ref{sec:method:indexing} ($p{=}0.9$, 40\% per-sentence keyword cap). Table~\ref{tab:trie-stats} and Figure~\ref{fig:trie-scaling} report means over the 30 inputs per length. Because the trie is constructed once per document and is invariant to the compression budget and query mode, these statistics hold unchanged across budget sweeps and multi-turn use.

Growth is in width, not depth. Path depth is scale-invariant: the median
stays between 2.1 and 2.4 and the maximum below 8.5 at every length, since
depth is bounded by the number of salient keywords per sentence,
$|K_i \cap \mathcal{S}|$, rather than by document length, so per-sentence
traversal cost does not grow with the input. Node count instead tracks the
number of sentences near-linearly: an $8\times$ increase in raw length
yields $8.6\times$ more sentences and $9.5\times$ more nodes (1.21 to 1.34
nodes per sentence), the mild rise being consistent with a larger salience
vocabulary reducing prefix sharing between sentences. The theme vocabulary
itself grows sublinearly: depth-1 branches and the salience set expand only
$6.7\times$ and $6.6\times$ over the same range, reflecting the bound
imposed by the salience quantile. In absolute terms the structure stays
small, at ${\approx}20.6$k nodes for a 256k-token document.

A stable $8$--$11\%$ of sentences contain no salience-set keyword and
attach at the root without a theme path (Table~\ref{tab:trie-stats}, last
row).
The dip at 64k (node ratio $1.64\times$, lower maximum depth) reflects
book-mix variance in that stretch of the stream rather than a scaling
effect, visible in the per-input rows.

\section{Accuracy}
\label{app:accuracy}

We extend our accuracy evaluation to Ministral-8B-Instruct to assess generalization beyond the models reported in the main paper. Table~\ref{tab:longbench_20_ministral} presents LongBench results across all task categories at a 20\% retention budget. Tables~\ref{tab:longbench_sweep_llama} and~\ref{tab:longbench_sweep_ministral} present full sweep results on LongBench.

\paragraph{Method Configurations}
All methods are implemented in PyTorch \citep{NEURIPS2019_torch} 2.6.0 or 2.7.1, depending on each method's release requirements and compatibility with FlashAttention2. H2O uses chunked prefill at 8k to avoid OOM on long prompts. CPC replaces its LLMLingua GPT-3.5 evaluator with a local Llama-3.1-8B-Instruct pipeline so all methods share the same answer generator. Sentinel adds token-aware chunking at sentence boundaries to the preprocessing script, as the original implementation feeds the full context to the proxy in one shot and OOMs on long LongBench inputs.

\begin{table*}[th]
\centering
\small
\caption{LongBench results on Ministral-8B-Instruct (20\% retention budget).}
\label{tab:longbench_20_ministral}
\begin{tabular}{l *{7}{c}}
\toprule
\textbf{Method}
& \textbf{Single-Doc QA}
& \textbf{Multi-Doc QA}
& \textbf{Summarization}
& \textbf{Few-Shot}
& \textbf{Synthetic}
& \textbf{Code}
& \textbf{Avg.} \\
\midrule
\multicolumn{8}{c}{\textbf{Ministral-8B-Instruct}} \\
\midrule
Full-context
& 41.66 & 49.45 & 27.73 & 71.00 & 54.50 & 66.30 & 51.77 \\
\midrule
\multicolumn{8}{c}{\textbf{KV Cache Methods (20\%)}} \\
\midrule
SnapKV        & 41.24 & 49.59 & 25.35 & 70.83 & 54.75 & 57.72 & 49.12 \\
FastKV        & 41.80 & 49.54 & 25.11 & 70.88 & 54.75 & 57.44 & 49.15 \\
DuoAttention  & 28.85 & 33.25 & 21.04 & 63.52 & 14.50 & 51.88 & 35.80 \\
SentenceKV    & 35.40 & 26.73 & 20.29 & 54.13 & 43.50 & 43.75 & 36.51 \\
\midrule
\multicolumn{8}{c}{\textbf{Preprocessing Methods (20\%)}} \\
\midrule
LLMLingua-2   & 29.98 & 32.45 & 23.21 & 41.71 & 11.75 & 34.52 & 29.66 \\
EXIT          & 32.71 & 32.23 & 24.01 & 65.30 & 13.25 & 40.01 & 35.58 \\
RECOMP        & 35.07 & 50.10 & 23.19 & 54.69 & 52.25 & 40.97 & 42.23 \\
CPC           & 39.05 & 43.99 & 24.17 & 51.78 & 50.25 & 11.50 & 37.53 \\
Sentinel      & 39.85 & 48.29 & 25.35 & 61.89 & 50.50 & 43.86 & 44.68 \\
\rowcolor{blue!10}
SALT          & 37.95 & 49.55 & 25.39 & 66.41 & 56.00 & 48.52 & 46.68 \\
\bottomrule
\end{tabular}
\end{table*}

\subsection{Multi-Turn Evaluation on QuALITY}
\label{sec:multiturn-quality}
We evaluate on the QuALITY \citep{pang2022quality} dev split (v1.0.1, html-stripped): multiple-choice
QA over long narrative documents, using a 50-article subset (972 question
turns). Each question is issued as a single conversational turn and scored
by \texttt{argmax} over the option-letter logits. All runs use
\textsc{Llama-3.1-8B-Instruct} in bf16 with scaled dot-product attention (SDPA) on a single H100, with a 10-token decode budget for time per output token (TPOT) parity.

We compare four configurations at a 20\% compression budget where
applicable: an uncompressed baseline; FastKV with $0.20$ KV-cache retain
rate; SALT with a per-article index built once and per-turn
theme-conditioned retrieval; and RECOMP-extractive with the published
\texttt{fangyuan/nq\_extractive\_compressor}, re-embedding sentences per
query. RECOMP's encoder is run in bf16 under \texttt{torch.autocast}
(embeddings cast to fp32 for scoring) for a $3\times$ speedup with no
measurable accuracy change. SALT and RECOMP are matched on input-token
budget ($\sim$1.1--1.2k tokens); FastKV leaves the prompt unchanged at
$\sim$5.7k tokens.

\begin{table}[t]
\centering
\small
\setlength{\tabcolsep}{3pt}
\caption{QuALITY @ 20\% budget. 50 articles, 972 turns.
Comp.\ = per-turn compression (ms). $\Sigma_{19}$ = 19-turn
cumulative cost (s).}
\label{tab:quality-20}
\begin{tabular}{lrrrrr}
\toprule
Method & TTFT & Comp. & Acc. & HARD & $\Sigma_{19}$ \\
       & (ms) & (ms)  &      &      & (s)            \\
\midrule
No compression & 195 & --- & 0.741 & 0.668 & 3.71 \\
FastKV         & 138 & --- & 0.739 & 0.664 & 2.62 \\
RECOMP         &  53 &  46 & 0.615 & 0.543 & 1.88 \\
\rowcolor{blue!10}
SALT           &  38 &  11 & 0.726 & 0.657 & 0.91 \\
\bottomrule
\end{tabular}
\end{table}

At matched budget, SALT essentially preserves baseline accuracy
(72.6\% vs.\ 74.1\% baseline, 73.9\% FastKV; $\Delta < 1.5$ pp) and
outperforms RECOMP-extractive by $\sim$11 points (72.6\% vs.\ 61.5\%).
Per-turn cost differs sharply: RECOMP re-encodes the document on every
query at 46 ms/turn, while SALT amortizes a 190 ms index across the
conversation and pays only $\sim$11 ms/turn thereafter. Over a 19-turn
dialogue, total compression+prefill cost is 3.71 s (baseline), 2.62 s
(FastKV), 1.88 s (RECOMP), and 0.91 s (SALT), a $4\times$ end-to-end
speedup over the baseline at near-equal accuracy, and a $2\times$ speedup
over RECOMP at much higher accuracy.

\section{Efficiency}
\label{app:efficiency}

We report the detailed efficiency measurements underlying the main results. Tables~\ref{tab:walltime-main} and~\ref{tab:memory-main-full} give end-to-end walltime and peak GPU memory, respectively, across context lengths from 16k to 256k tokens.

\begin{table}[t]
\centering
\small
\setlength{\tabcolsep}{6pt}
\caption{Walltime (s) across context lengths. Methods are grouped into preprocessing (top) and KV-cache (bottom).}
\label{tab:walltime-main}
\begin{tabular}{lccccc}
\toprule
Method & 16k & 32k & 64k & 128k & 256k \\
\midrule
\rowcolor{blue!10}SALT          & \textbf{1.92}  & \textbf{2.61}  & 4.01           & \textbf{5.99}  & \textbf{11.51} \\
EXIT          & 5.83           & 16.30          & 60.60          & OOM            & OOM            \\
CPC           & 3.24           & 5.44           & 10.13          & 21.67          & 42.25          \\
RECOMP        & 2.00           & 2.29           & 3.34           & 6.27           & 11.95          \\
Sentinel       & 2.32           & 2.81           & 4.02           & 6.49           & 12.03          \\
\midrule
FastKV        & 1.73           & 2.35           & 4.22           & 10.77          & 34.29          \\
SnapKV        & 2.00           & 3.08           & 6.43           & 18.13          & 61.75          \\
DuoAttention  & 2.50           & 3.24           & 5.32           & 11.72          & 33.40          \\
SentenceKV     & 10.43          & 17.06          & 31.20          & 63.80          & 64.40          \\
\bottomrule
\end{tabular}
\end{table}

\begin{figure*}[t]
  \centering
  \includegraphics[width=\textwidth]{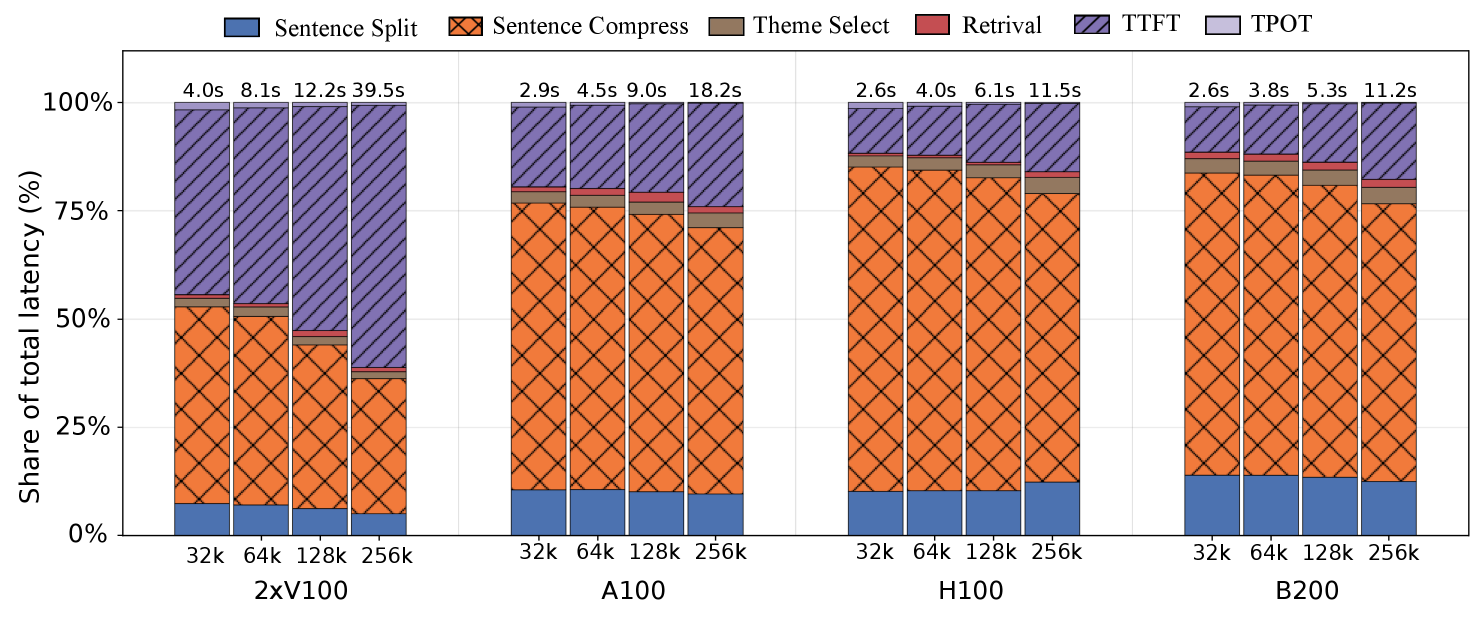}
  \caption{End-to-end latency and TPOT of SALT on Llama-3.1-8B-Instruct across different NVIDIA GPUs and raw-context lengths at a 20\% retention budget, normalized so each bar sums to 100\%.}
  \label{fig:hardware-portability-latency-normalized}
\end{figure*}

\section{Hardware}
\label{app:hardware_extra}

Normalizing each run to 100\% exposes a distinct shift in the component breakdown that is obscured by looking at absolute latency alone. For Ampere, Hopper, and Blackwell architectures, TTFT consistently accounts for roughly a third of the total time, while preprocessing forms the next largest segment and per-token decode fills the remainder. 

The legacy V100 hardware breaks this pattern because TTFT becomes the dominant non-compression component, outlasting preprocessing at every context length. This divergence is likely architectural rather than algorithmic. The Volta generation lacks native bf16 support and cannot utilize the FlashAttention kernels designed for the tensor cores found in Ampere and subsequent architectures. As a result, the post-compression prefill attention must run in fp16 along an unoptimized execution path. Conversely, the compression phase is heavily dominated by lighter scoring and selection operations, making it relatively unaffected by these hardware limitations. This explains why the compression phase occupies a larger percentage share on newer hardware even as its absolute processing time decreases.

\begin{table*}[th]
\centering
\small
\caption{LongBench average against retention budget on LLaMA-3.1-8B-Instruct.}
\label{tab:longbench_sweep_llama}
\begin{tabularx}{\textwidth}{l *{9}{>{\centering\arraybackslash}X}}
\toprule
\textbf{Method} & \textbf{10\%} & \textbf{20\%} & \textbf{30\%} & \textbf{40\%} & \textbf{50\%} & \textbf{60\%} & \textbf{70\%} & \textbf{80\%} & \textbf{90\%} \\
\midrule
\multicolumn{10}{c}{\textbf{KV Cache Methods}} \\
\midrule
SnapKV         & 46.52 & 47.25 & 47.56 & 47.72 & 47.96 & 48.16 & 48.09 & 48.20 & 48.28 \\
FastKV         & 46.80 & 47.54 & 47.84 & 47.92 & 48.04 & 48.10 & 48.17 & 48.23 & 48.30 \\
DuoAttention   & 27.99 & 41.96 & 45.20 & 46.31 & 47.41 & 47.51 & 47.72 & 47.84 & 48.09 \\
SentenceKV     & 46.56 & 46.42 & 44.79 & 46.88 & 46.91 & 46.90 & 46.91 & 46.85 & 46.92 \\
\midrule
\multicolumn{10}{c}{\textbf{Preprocessing Methods}} \\
\midrule
LLMLingua-2    & 23.34 & 26.02 & 29.11 & 32.09 & 37.90 & 43.25 & 45.43 & 46.52 & 46.36 \\
EXIT           & 29.27 & 32.30 & 34.68 & 36.90 & 38.97 & 40.87 & 42.87 & 44.87 & 46.87 \\
RECOMP         & 37.08 & 39.98 & 41.80 & 42.28 & 43.54 & 44.41 & 45.04 & 45.34 & 45.67 \\
CPC            & 35.82 & 37.74 & 39.41 & 39.80 & 40.38 & 40.60 & 40.95 & 39.94 & 40.03 \\
Sentinel       & 40.03 & 39.47 & 40.65 & 42.01 & 43.69 & 43.81 & 45.26 & 45.37 & 46.65 \\
\rowcolor{blue!10}
SALT           & 39.58 & 43.51 & 43.68 & 44.44 & 45.83 & 46.18 & 47.25 & 47.42 & 47.27 \\
\bottomrule
\end{tabularx}
\end{table*}

\begin{table*}[th]
\centering
\small
\caption{LongBench average against retention budget on Ministral-8B-Instruct.}
\label{tab:longbench_sweep_ministral}
\begin{tabularx}{\textwidth}{l *{9}{>{\centering\arraybackslash}X}}
\toprule
\textbf{Method} & \textbf{10\%} & \textbf{20\%} & \textbf{30\%} & \textbf{40\%} & \textbf{50\%} & \textbf{60\%} & \textbf{70\%} & \textbf{80\%} & \textbf{90\%} \\
\midrule
\multicolumn{10}{c}{\textbf{KV Cache Methods}} \\
\midrule
SnapKV         & 48.49 & 49.12 & 49.29 & 49.45 & 49.50 & 49.72 & 49.75 & 49.72 & 49.75 \\
FastKV         & 48.42 & 49.15 & 49.42 & 49.57 & 49.78 & 49.81 & 49.79 & 49.77 & 49.83 \\
DuoAttention   & 32.22 & 35.80 & 39.38 & 47.40 & 49.03 & 49.11 & 49.11 & 49.34 & 49.54 \\
SentenceKV     & 32.86 & 36.51 & 36.51 & 39.60 & 41.68 & 44.72 & 45.72 & 47.74 & 48.94 \\
\midrule
\multicolumn{10}{c}{\textbf{Preprocessing Methods}} \\
\midrule
LLMLingua-2    & 25.74 & 29.66 & 32.52 & 36.21 & 40.37 & 44.64 & 47.64 & 48.31 & 48.26 \\
EXIT           & 32.17 & 35.58 & 38.02 & 39.86 & 40.79 & 43.09 & 44.15 & 45.77 & 46.99 \\
RECOMP         & 38.86 & 42.23 & 43.92 & 45.65 & 46.77 & 47.37 & 48.01 & 47.99 & 48.38 \\
CPC            & 35.58 & 37.53 & 38.78 & 42.09 & 43.15 & 45.29 & 45.93 & 47.78 & 48.39 \\
Sentinel       & 42.25 & 44.68 & 46.35 & 46.88 & 47.02 & 47.55 & 47.85 & 48.02 & 47.96 \\
\rowcolor{blue!10}
SALT           & 44.78 & 46.68 & 47.35 & 47.52 & 47.92 & 48.15 & 48.83 & 48.93 & 48.87 \\
\bottomrule
\end{tabularx}
\end{table*}

\begin{table}[t]
\centering
\small
\setlength{\tabcolsep}{6pt}
\caption{Peak GPU memory (GB) across context lengths. Methods are grouped into preprocessing (top) and KV-cache (bottom).}
\label{tab:memory-main-full}
\begin{tabular}{lccccc}
\toprule
Method & 16k & 32k & 64k & 128k & 256k \\
\midrule
\rowcolor{blue!10}SALT          & \textbf{16.70} & \textbf{17.90} & \textbf{20.20} & \textbf{23.30} & \textbf{32.80} \\
EXIT          & 23.70          & 45.40          & 73.20          & OOM            & OOM            \\
CPC           & 20.49          & 23.13          & 28.46          & 38.00          & 58.51          \\
RECOMP        & 18.50          & 21.30          & 27.20          & 37.70          & 60.40          \\
Sentinel       & 19.10          & 21.40          & 26.30          & 33.90          & 54.10          \\
\midrule
FastKV        & 17.37          & 19.39          & 23.43          & 31.52          & 47.67          \\
SnapKV        & 17.57          & 19.79          & 24.20          & 33.12          & 50.88          \\
DuoAttention  & 16.80          & 17.85          & 19.85          & 23.85          & 31.86          \\
SentenceKV     & 22.98          & 28.50          & 39.80          & 60.73          & 60.89          \\
\bottomrule
\end{tabular}
\end{table}

\end{document}